\documentclass{emulateapj} 

\usepackage{graphics} 

\shorttitle{The torus in NGC\,4151.}
\shortauthors{Riffel, Storchi-Bergmann \& McGregor}

\begin{document}


\title{The Dusty Nuclear Torus in NGC\,4151: Constraints from Gemini Near-Infrared Integral Field Spectrograph Observations }


\author{Rogemar A. Riffel, Thaisa Storchi-Bergmann 
}
\affil{Universidade Federal do Rio Grande do Sul, IF, CP 15051, Porto
Alegre 91501-970, RS, Brazil} \email{rogemar@ufrgs.br}

\and

\author{Peter J. McGregor
} \affil{Research School of Astronomy and
Astrophysics, Australian National University, Cotter Road, Weston Creek,
ACT 2611, Australia}






\begin{abstract} 
We have used a near-infrared nuclear spectrum (covering the Z, J, H and K bands) of the nucleus of NGC\,4151 obtained with the Gemini Near-infrared Integral Field Spectrograph (NIFS) and adaptive optics, to isolate and constrain the properties of a near-IR unresolved nuclear source whose spectral signature is clearly present in our data. The near-IR spectrum was combined with an optical spectrum obtained with the  Space Telescope Imaging Spectrograph which was used to constrain the contribution of a power-law component. After subtraction of the power-law component,
the near-IR continuum is well fitted by a blackbody function, with $T=1285\pm50\,$K, which dominates the nuclear spectrum -- within an aperture of radius 0$\farcs$3 -- in the near-IR. We attribute the blackbody component to emission by a dusty structure, with hot dust mass $M_{\rm HD}=(6.9\pm\,1.5)\,\times10^{-4}\,{\rm M_\odot}$, not resolved by our observations, which provide only an upper limit for its distance from the nucleus of 4\,pc.  If the reddening derived for the narrow-line region also applies to the near-IR source, we obtain a temperature $T=1360\pm50\,$K and a  mass $M_{\rm HD}=(3.1\pm\,0.7)\,\times10^{-4}\,{\rm M_\odot}$ for the hot dust.
This  structure may be the inner wall of the dusty torus postulated by the Unified Model or the inner part of a dusty wind originating in the accretion disk.
\end{abstract}





\section{Introduction}

The presence of a very compact infrared source in the nucleus of  the Seyfert galaxy NGC\,4151 was first suggested by the observations of  \citet{penston74}, who reported photometric variations of the galaxy nucleus both in the optical and in the near-infrared (hereafter near-IR).  A close inspection of the light curves shown by  \citet{penston74} suggests a delay between the near-IR and optical variations, recently confirmed by \citet{minezaki04}, who could better quantify its value as 48$^{+2}_{-3}$ days. These authors concluded that the $K-$band emission is dominated by thermal radiation from hot dust located at $\approx$0.04\,pc from the nucleus. A similar intepretation was previously proposed by \citet{rieke81}, via models fitted to a large aperture ($\sim$\,8\arcsec) nuclear optical and near-IR spectrum, concluding that in the  near-IR the emission is dominated by thermal reradiation by hot dust with temperature $T_{\rm HD}=1300-1500$\,K. 

\citet{mundell03} have reported neutral hydrogen absorption in VLA radio observations  with subparsec linear resolution, close to the systemic  velocity of the galaxy, interpreted as due to a clumpy gas layer suggested to be associated to a circumnuclear obscuring torus. According to \citet{mundell03}, this absorption is observed along the line of sight towards the first component of the radio counterjet, at $\le$0.1\,pc from the nucleus, thus consistent with the estimated inner radius of $\approx$0.04\,pc derived by \citet{minezaki04}. The interpretation put forth by \citet{mundell03} -- that the obscuring torus is not in front of the active galactic nucleus (AGN) -- is necessary in order to reconcile the radio observations with the UV/optical  observations which show a blue nuclear continuum and broad emission lines, indicating that both the nuclear source and the broad-line region are visible.

The presence of a dusty torus around the nucleus of NGC\,4151 was disputed by \citet{swain03}  on the basis of interferometric observations at 2.2$\mu$m using the two 10\,m Keck telescopes. These authors found a marginally resolved nuclear source $\le$0.1\,pc in diameter -- consistent with the observations of both \citet{mundell03} and \citet{minezaki04}, but  favor an origin for this emission in thermal gas from the accretion disk, arguing that  previous nuclear spectra of NGC\,4151 did not show an  ``infrared bump'', in apparent contradiction with the previous study by  \citet{rieke81}.  Such a bump is expected if a dusty torus is present, due to reprocessing of UV/optical radiation from the central engine \citep[e.g.][]{rieke81,barvainis87,sanders89}.

The alternative interpretation proposed by \citet{swain03} seems nevertheless to be in line with recent results presented by \citet{kraemer08}, who studied the distribution of [O{\sc\,iii}]$\lambda$5007 and [O{\sc\,ii}]$\lambda$3727 emission in the Narrow Line Region (NLR) of NGC\,4151. These authors pointed out that there does not seem to be a region of complete shadow around the apex of the bicone, as significant ionized gas emission is found there, although with a lower ionization parameter, consistent with a weaker ionizing flux. \citet{kraemer08}  propose that the attenuation of the nuclear radiation could be due to low-ionization absorbers swept out by a wind originating in the accretion disk, suggesting that there is no need for a ``classical, doughnut-shaped'' torus in order to explain the observed light distributions, in agreement with recent modelling for the obscuring region around an AGN \citep{elitzur08}.

In this  work, we present new observations of the nucleus of NGC\,4151 with the Gemini Near infrared Integral Field Spectrograph (NIFS) at an angular resolution of 0$\farcs$12 -- corresponding to a spatial resolution of a few pc at the galaxy -- which is much improved relative to previous spectroscopic studies. Our observations clearly reveal the presence of an unresolved nuclear source in the near-IR. We have already used the NIFS data
to map the excitation \citep[][hereafter Paper I]{sb09}  and kinematics \citep{sl09} of the NLR. In the present paper, we use the nuclear spectrum, combined with surrounding extranuclear spectra and a {\it Space Telescope Imaging Spectrograph} (STIS) optical spectrum, in order to isolate the emission of the infrared nuclear source, which is consistent with that predicted for
the torus postulated in the Unified Model of AGNs \citep{antonucci93}.

We adopt a distance to NGC\,4151 of 13.3\,Mpc, corresponding to a scale at the galaxy of 65\,pc\,arcsec$^{-1}$ \citep{mundell03}. This paper is organized as follows. In section\,2, we present a summary of the observations, in section\,3 we present the results and in section\,4 we present and discuss our conclusions.




\section{Observations}

Two-dimensional spectroscopic data were obtained on the Gemini North telescope with the instrument NIFS \citep{nifs03} operating with the adaptive optics module ALTAIR on the nights of December 12 and 13, 2006.  
The observations covered the spectral bands Z, J, H and K, at  spectral resolutions of 4990, 6040, 5290 and 5290, respectively, resulting in a wavelength coverage from 0.94$\mu$m to 2.40$\mu$m. More details of the observations and reduction procedures can be found in \citet{sb09}.

\begin{figure}
\centering
\includegraphics[scale=0.5]{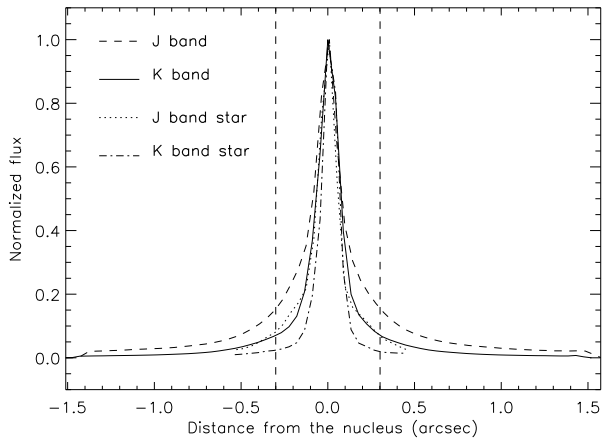}
\caption{Spatial profiles for the galaxy and of a star at J and K bands. The vertical dashed lines mark the extraction aperture of the nuclear spectrum (0\farcs3 radius).} 
\label{psf}
\end{figure}

In the present paper  we concentrate on the analysis of the nuclear continuum. The spatial profiles of the nuclear region of the galaxy  in the K and J bands are shown in Fig.\ref{psf}, in comparison with stellar profiles. The nuclear source -- which has a full width at half maximuum (FWHM) of 0\farcs12$\pm$0\farcs02  in the K-band and 0\farcs16$\pm$0\farcs02 in the J-band -- corresponding to 8.0$\pm$1.3\,pc and 10.4$\pm$1.3\,pc at the galaxy, respectively -- is unresolved by our observations, in spite of the fact that their spatial profiles are marginally broader than those of the stars in Fig.\,\ref{psf}. As discussed in \citet{sb09}, this is due to the much shorter exposures of the stellar observations. 

In order to include as much as possible of the nuclear source, we have 
chosen an extraction aperture of 0\farcs3 radius which includes the nuclear flux down to 10\,\% of the peak value in the K-band, as illustrated by the dashed lines in Fig.\,\ref{psf}.

\begin{figure}
\centering
\includegraphics[scale=0.52]{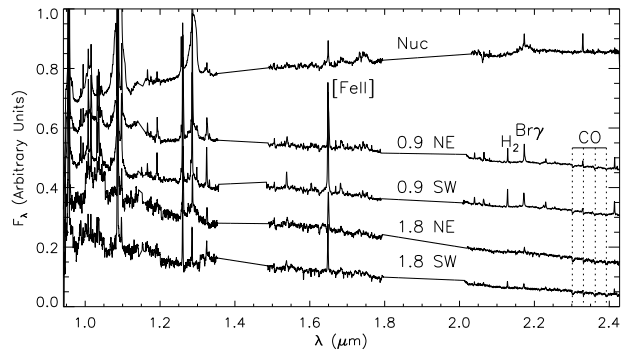}
\caption{Nuclear spectrum (top) obtained for a circular aperture with 0\farcs3 radius and extra-nuclear spectra at 
the positions 0\farcs9\,NE, 0\farcs9\,SW, 1\farcs8\,NE and 1\farcs8\,SW, normalized to the flux at the nucleus at $\lambda\approx\,1\mu$m.The spectra are shown in arbitrary flux units and are shifted by a constant for clarity.} 
\label{spec}
\end{figure}

\section{Results} 

The nuclear spectrum is shown in the Fig.\ref{spec}, together with extranuclear spectra obtained through similar 0$\farcs$3 apertures at distances of 0\farcs9\,NE, 0\farcs9\,SW, 1\farcs8\,NE, and  1\farcs8\,SW from the nucleus, normalized to the flux of the nuclear spectrum at $\lambda\approx1\mu$m. Fluxes of more than 50 emission lines have been  measured and listed in Table 1 of Paper I for  the nuclear and two extranuclear spectra.

From Fig.\,\ref{spec}, as well as from spectra shown in Paper I, we observe that the extranuclear spectra  are blue at all locations around the nucleus. As this blue continuum is observed up to at least 2\,arcsec from the nucleus, it cannot be due to the AGN and can thus be attributed to a young stellar population surrounding the nucleus. Absorption features from stellar CO bands are indeed  observed in the K band,  marked by vertical dotted lines in Fig.\,\ref{spec}. Only right at the nucleus the continuum is clearly very red. Similar red nuclear continua have been observed for other AGNs and attributed to emission from hot dust, possibly from a dusty torus around the supermassive black hole, with temperature close to the dust sublimation temperature  \citep{barvainis87,marco98,marco00,ardila05b,ardila06,riffel09}.

\subsection{Nuclear optical continuum}

\begin{figure*}
\centering
\includegraphics[scale=0.8]{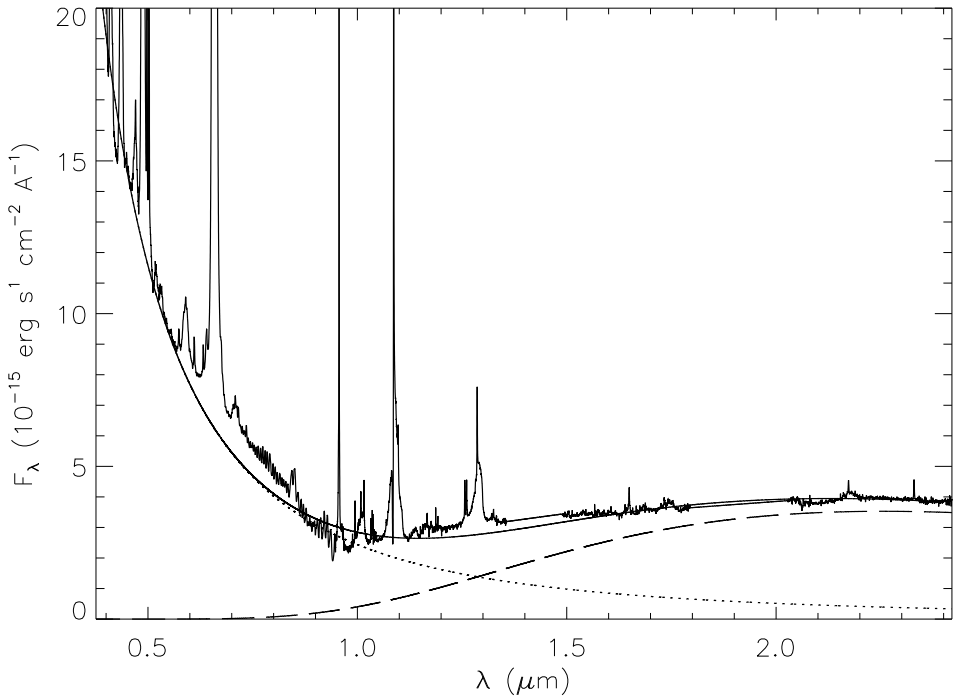}
\caption{The optical$+$near-IR nuclear spectrum of NGC\,4151, together with a fit (continous line) considering the contribution of a power-law (dotted line) plus a black body function (dashed line).} 
\label{fit}
\end{figure*}

In order to better constrain the nuclear spectral energy distribution (SED), we have used an optical STIS spectrum of the nucleus extracted within a similar aperture to that of our near-IR spectrum, obtained from  the  atlas of AGN spectra compiled by \citet{spinelli06}. This spectrum was extracted within an aperture of  0\farcs2$\times$0\farcs1 and covers the spectral range 0.3--1\,$\mu$m. We have used the continuum region from 9780 to 9820\,\AA~ to normalize the optical spectrum to the same flux as our near-IR spectrum (which extends down to the Z-band) as there is a flux difference of $\approx$\,30\% between the STIS spectrum and our nuclear spectrum. The small difference between the apertures is not important because the stellar contribution, when compared to the nuclear one, is small, as discussed below.  We show the optical plus near-IR spectrum in Fig.\,\ref{fit}. The optical continuum is very blue and can be well  reproduced by a power-law 
up to $\approx$\,1\,$\mu$m, where there is a change in slope and  a red component begins to appear in the near-IR. Although the near-IR emission is dominated by this component, the contribution of the  power-law in the near-IR is not negligible in the Z, J and H bands, as discussed below.

\subsection{Modelling the spectral energy distribution}

We have modelled the optical$+$near-IR nuclear continuum of NGC\,4151 by the sum of two components: a power law, which seems to dominate in the optical, plus a blackbody function, which dominates in the near-IR, as illustrated in Fig.\,\ref{fit}. 
The fit was performed using the {\it nfit1d} task from the {\sc stsdas}  {\sc iraf} package, which allows the selection of limited spectral intervals for the fit. We have excluded the spectral interval between 0.6 and 0.9$\mu$m from the fit as we have estimated that the contribution of the Paschen continuum in this interval ($\approx\,1.3\times10^{-15}$\,ergs\,cm$^{-2}$s$^{-1}$A$^{-1}$ at 8200\AA), is not negligible when compared to the total flux observed in the continuum. Moreover, the quality of the optical spectrum in its red end is quite poor, and should not be used to constrain the SED. Excluding the above spectral region, the  best fit was obtained for a power law $F_\lambda\propto\lambda^{-2.25}$  and a blackbody function of temperature $T=1285\pm50\,$K, which  peaks at $\approx2.3\,\mu$m. The fit of the two functions combined is shown in Fig.\,\ref{fit} as a continuous line, while the black body component is shown as a dashed line and the power law as dotted line.

We have next evaluated the effect  of the underlying stellar population in  the fit above. A stellar population spectrum was extracted from our near-IR datacubes within a ring around the nucleus with inner radius of 0\farcs6 and outer radius of 0\farcs9. We have then normalized the flux of this extranuclear spectrum to 15\% of the peak nuclear flux in the K band, which is the maximum contribution of the stellar population estimated from a spatial profile through the nucleus. Under the assumption that the underlying stellar population in the nuclear spectrum is the same as the one from the ring (which is supported by other spectra from the nuclear vicinity), we have then subtracted the extranuclear spectrum from the nuclear one. As we do not have an extranuclear optical spectrum, we repeated the fit for the near-IR only, comparing the temperature of the uncorrected to the corrected spectrum, concluding that, within the uncertainties ($\approx$50\,K), the blackbody temperature is the same. 

The reddening may also affect the results of the fit. An estimate of the reddening in the near-IR can be obtained from the emission lines of the narrow-line region. We have used the  [Fe\,{\sc ii}]$\lambda1.2570\mu$m/$\lambda1.6440\mu$m line ratio to calculate the reddening (as described in Paper I, but using the present nuclear spectrum, for an aperture of radius 0$\farcs$3), resulting in  E(B-V)$\approx0.6$. It is not clear that this reddening applies to the optical continuum. If it does, it would result in a very steep power law for the corrected nuclear continuum: using the extinction curve of \citet{rieke85}, we obtain $F_\lambda\propto\lambda^{-4.5}$. Nevertheless, the presence of broad lines in the spectra and the already very blue continuum suggests that the reddening of the optical spectrum is much smaller; we may be seing the nuclear source through a low extinction window to the nucleus. This may apply also to the near-IR nuclear source, but, even if the extinction applies, the correction in the near-IR is small. A reddening correction of only the near-IR spectrum for  E(B-V)\,=\,0.6, keeping the power-law index the same, results in a revised blackbody temperature of $T=1360\pm50\,$K. 

\subsection{The hot dust mass}

Following \citet{barvainis87} and \citet{riffel09} we can estimate the mass of the hot dust producing the unresolved near infrared emission.  Firstly, we obtain the infrared spectral luminosity of each dust grain, in erg\,s$^{-1}$\,Hz$^{-1}$, assuming that the dust is composed of graphite grains, by
\begin{equation}
 L^{\rm gr}_{\rm \nu,ir} = 4\pi^2a^2 Q_\nu B_\nu(T_{\rm gr}),
\label{lgr}
\end{equation}
where $a$ is the grain radius, $Q_\nu$ is the absorption efficiency and $B_\nu(T_{\rm gr})$ is its spectral distribution assumed to be a Planck function with temperature $T_{\rm gr}$. 

The number of dust grains ($N_{\rm HD}$) can be estimated from the ratio between the total luminosity of the hot dust ($L_{\rm ir}^{\rm HD}$) and the total luminosity of one dust grain  ($L_{\rm ir}^{\rm gr}$) as $N_{\rm HD}= \frac{L_{\rm ir}^{\rm HD}}{L_{\rm ir}^{\rm gr}}$. 
$L_{\rm ir}^{\rm HD}$ was obtained by integrating the flux under the Planck function fitted to the nuclear spectrum  and adopting a distance to NGC\,4151 of $d=13.3\,$Mpc, while $L_{\rm ir}^{\rm gr}$ was obtained by the intregration of equation\,\ref{lgr} for a temperature of $T_{gr}=1285\,$K.
Finally, the mass of the emitting hot dust is obtained by:

\begin{equation} 
M_{\rm HD}\approx\frac{4\pi}{3}a^3 N_{\rm HD}\rho_{\rm gr}.
\end{equation}

Assuming a graphite density of $\rho_{\rm gr}=2.26\,$g\,cm$^{-3}$ \citep{granato94}, we obtain $M_{\rm HD}=(6.9\pm\,1.5)\times10^{-4}\,{\rm M_\odot}$. 

If we assume that the near-IR nuclear source is subject to reddeninng, we should then use the temperature $T_{gr}=1360\,$K, which will lead to a smaller dust mass of  $M_{\rm HD}=(3.1\pm\,0.7)\,\times10^{-4}\,{\rm M_\odot}$.

\begin{table*}
\centering
\caption{Masses of hot dust found in AGNs}
\vspace{0.3cm}
\begin{tabular}{l l l}
\hline
 Galaxy                & $M_{\rm HD}$ (M$_\odot$)& Reference \\
\hline
NGC\,4151 &              $6.9\times10^{-4}$    & This work (no extinction, see text)                      \\
NGC\,7582 &              $2.8\times10^{-3}$          &  \citet{riffel09}                       \\
Mrk\,1239 &              $2.7\times10^{-2}$          &  \citet{ardila06}           \\
Mrk\,766   &             $2.1\times10^{-3}$          & \citet{ardila05b}          \\
NGC\,1068 &              $1.1\times10^{-3}$          &  \citet{marco00}                     \\
NGC\,7469 &              $5.2\times10^{-2}$          & \citet{marco98}                      \\
NGC\,4593  &             $5.0\times10^{-4}$          & \citet{santos95}                      \\ 
NGC\,3783  &             $2.5\times10^{-3}$          & \citet{glass92}                      \\
NGC\,1566  &             $7.0\times10^{-4}$          &  \citet{baribaud92}                  \\
Fairall\,9 &             $2.0\times10^{-2}$          & \citet{clavel89}                     \\
\hline
\end{tabular}
\label{hd_mass}
\end{table*}

In table\,\ref{hd_mass}, we present a comparison of the mass of hot dust obtained here for NGC\,4151 with the ones derived for other AGNs in previous works. This comparison reveals that the hot dust mass for NGC\,4151 is similar to the ones obtained for NGC\,1068 by \citet{marco00}, for NGC\,4593 by \citet{santos95} and  for NGC\,1566 by \citet{baribaud92} and is smaller than the masses derived for other galaxies such as NGC\,7582, Mrk\,1239, Mrk\,766, NGC\,7469, NGC\,3783 and Fairall\,9 by \citet{riffel09}, \citet{ardila06}, \citet{ardila05b},\citet{marco98}, \citet{glass92} and  \citet{clavel89}, respectively.

\section{Discussion and Conclusions}

As discussed in the Introduction, \citet{swain03} observed NGC\,4151 at 2.2\,$\mu$m and have marginally resolved a nuclear source  $\le$0.1\,pc in diameter, concluding that the emission was due to thermal gas from the central accretion disk instead of a dusty torus, as they claimed that previous observations did not show the bump. Our near-IR nuclear spectrum (Fig.\,\ref{spec}) -- clearly shows this bump and favours emission by hot dust as the dominant mechanism for its origin, in agreement with previous studies using larger aperture data \citep{rieke81}. A more recent near-IR spectrum obtained by \citet{riffel06}, although with lower spatial resolution than ours, also suggests the presence of the bump. 

The modern view of the torus \citep{elitzur08} is not of a static doughnut shape structure but of a clumpy wind of dusty and optically thick clouds originating in the black-hole accretion disk, a view which has recently been favored also in the case of NGC\,4151 by \citet{kraemer08}, and which is in partial agreement with the proposition by \citet{swain03}. Being clumpy, the dusty wind allows the view of the nuclear source through low-extinction holes. The presence of such holes is also supported by the observation of ionized gas close to the apex of the bicone, beyond its presumed walls. This more updated view is also consistent with  our data and results from previous papers discussed above, if we consider that the near-IR nuclear source can be identified with the hottest part of the dusty wind, which is the one closest to the nucleus and which emits most of the near-IR radiation. According to our data, the temperature of this dust is  $T=1285 \pm50\,$K and its mass is $M_{\rm HD}=(6.9\pm\,1.5)\,\times10^{-4}\,{\rm M_\odot}$, obtained by fitting the optical and near-IR spectrum. The observation of the nuclear continuum and of the broad emission lines suggest that we are looking to the nuclear source through a low extinction window, in spite of the fact that the estimated reddening from the emission lines of the narrow-line region is E(B-V)\,=0.6. If we assume that this reddening is also affecting the nuclear source, we obtain  $T=1360 \pm50\,$K  for the dust temperature and mass $M_{\rm HD}=(3.1\pm\,0.7)\,\times10^{-4}\,{\rm M_\odot}$. The temperatures obtained are in good agreement with those previously obtained by  \citet{rieke81}. Comparison between our near-IR flux values with theirs shows that the nucleus was 3--4 times brighter than in the epoch of our observations, suggesting that the temperature of the hot dust does not depend on the AGN luminosity.

Regarding the location of the hot dust, our spatial resolution allows us to put only an upper limit to the radius of the infrared source of 4\,pc (half the FWHM of the nuclear source). 
Similar scales have been probed by \citet{sb05}, who found a heavily obscured starburst closer than 9\,pc from the nucleus in NGC\,1097 and argued that it could be located in the outskirts of the torus. \citet{jaffe04} using mid-IR interferomentric observations constrained the dust emission in NGC\,1068 to originate in a region of only 1\,pc diameter. In the case of NGC\,4151, optical and near-IR monitoring observations of \citet{minezaki04} provide a better constraint on the location of the inner radius of the torus at $\approx$0.04\,pc from the nucleus.

The NLR of NGC\,4151 has an approximate biconical morphology  with projected opening angle of $\sim$75$\degr$ and with its axis oriented along a PA $\sim$60/240$\degr$ (NE--SW). The SW side of the bicone is tilted toward us, making an angle of 45$\degr$ with the line-of-sight, according to \citet{das05}, who gives also an opening angle of the cone of 66$\degr$, which puts our line-of-sight just outside the cone wall. If the torus is oriented perpendicularly to the cone, we are looking down into the torus hole. This allows us to constrain the half-height of the torus to less than 0.04\,pc for a ``doughnut shape'' opaque torus. This torus should extend to at least 0.1\,pc from the nucleus, in order to contain the neutral hydrogen clouds observed by \citet{mundell03}. As proposed by these authors, the NE part of the radio jet would be behind the torus, while the SW part would be in front of it.

\acknowledgments We thank the referee Dr. George Rieke for valuable suggestions which helped to improve the present paper. This work is  based on observations obtained at the Gemini Observatory, which is operated by the
Association of Universities for Research in Astronomy, Inc., under a cooperative agreement
with the NSF on behalf of the Gemini partnership: the National Science Foundation (United
States), the Science and Technology Facilities Council (United Kingdom), the
National Research Council (Canada), CONICYT (Chile), the Australian Research Council
(Australia), Ministério da Ciência e Tecnologia (Brazil) 
and Ministerio de Ciencia, Tecnología e Innovación Productiva  (Argentina). This work has been partially supported by the Brazilian intitution CNPq and Australian Research Council.

{}   
\clearpage
\end{document}